\documentclass[aps,twocolumn,superscriptaddress,showpacs]{revtex4}
\usepackage{amsmath,amsthm,amssymb}

\usepackage{amsmath,graphicx}
\usepackage{color,ulem}
\newlength{\figwidth}
\newlength{\figlarge}
\begin{document}
\title{Correlation effects on a topological insulator at finite temperatures}

\author{Tsuneya Yoshida}
\author{Satoshi Fujimoto}
\author{Norio Kawakami}

\affiliation{
Department of Physics, Kyoto University, 
Kyoto 606-8502, Japan}

\date{\today}
\begin{abstract}
We analyze the effects of the local Coulomb interaction on a topological band insulator (TBI) by applying the dynamical mean field theory to a generalized Bernevig-Hughes-Zhang model having electron correlations. It is elucidated how the correlation effects modify electronic properties in the TBI phase at finite temperatures. 
In particular, the band inversion character of the TBI inevitably leads to the large reduction of
the spectral gap via the renormalization effect,
which results
in the strong temperature-dependence of the spin Hall conductivity. We clarify that a quantum phase transition from the TBI to a trivial Mott insulator, if it is nonmagnetic, is of first order with a hysteresis. This is confirmed via the interaction dependence of the double occupancy and the spectral function. A magnetic instability is also addressed.
All these results imply that
the spectral gap does not close at the transition. 
\end{abstract}
\maketitle

\section{Introduction}

Recently, the topological band insulator (TBI) has received much attention as a new research platform in condensed matter physics \cite{Hasan10,Qi10}.
The TBI has a non-trivial band structure resulting from spin-orbit interactions (SOIs). A remarkable property of this phase is that while it has the charge excitation gap in the bulk, an odd number of gapless edge states appear at boundaries \cite{Bernevig06}, which are robust against nonmagnetic impurities due to time reversal symmetry. The non-trivial band structure in the TBI was originally proposed for graphene \cite{Kane05_1,Kane05_2} and also for $\mathrm{HgTe/CdTe}$ quantum wells \cite{Bernevig06_BHZ}. The latter TBI was confirmed by transport measurements \cite{exp_2D-QW_MKonig}. Afterwards, TBIs in three dimensions were proposed  theoretically \cite{3DTI_Fu07,TBI_3DFu07,Qi08}, and confirmed in several bismuth-based compounds through the angle-resolved photo-emission spectroscopy measurements which elucidated the existence of the gapless edge states \cite{exp_3D-bismuth_DHsieh,exp_3D-bismuth_YXia,exp_3D-bismuth_YLChen}.

While the TBI can be described by a non-interacting band theory, correlation effects on this phase have attracted much attention in these days, since novel aspects of electron correlations would appear under topologically nontrivial conditions. This issue has further been stimulated by the fact that there are some iridium oxides, such as $\mathrm{Na_2IrO_3}$ \cite{NaIrO_Nagaosa09} and $\mathrm{A_2Ir_2O_7}$ $(\mathrm{A=Pr, Eu})$ \cite{TMI_LBalents09}, which could be candidates of TBIs with strong electron correlation. Theoretical studies on the correlation effects done thus far are roughly categorized into two types.  One is concerned with the case where the nonlocal repulsive interaction could induce a quantum phase transition, thus leading to topological insulators, as demonstrated theoretically for both 2D and 3D systems with mean field analysis \cite{Raghu08,TM-Z_YZhang09,Kurita11,Fiete11}. The other focuses on how the TBI phase is changed under strong electron correlations. For example, a competition between an antiferromagnetic (AF) phase and the TBI phase was studied by numerical approaches \cite{Yamaji11,AFvsTBI_DQMC_Hohenadler11,AFvsTBI_DQMC_edge_early_Zheng10,AFvsTBI_VCA_Yu11,AFvsTBI_CDMFTWu11} as well as mean field theory \cite{Rachel10}. According to a quantum Monte Carlo study by Hohenadler \textit{et al.} \cite{AFvsTBI_DQMC_Hohenadler11}, it was clarified that the TBI phase can change into a topologically trivial AF phase, which was supported by the bulk phase diagram obtained by a variational cluster approach and also a cluster dynamical mean field study \cite{AFvsTBI_VCA_Yu11,AFvsTBI_CDMFTWu11}. The former study also clarified the absence of edge states in the AF phase \cite{AFvsTBI_VCA_Yu11}. A competition of the topological phase and a charge-density-wave phase was also studied \cite{Varney10,Wang10}.  According to the exact diagonalization analysis, the nearest neighbor interaction induces a first-order transition, and the TBI phase changes into the charge-density-wave phase.

In spite of these intensive studies, it is still not clear how the electron correlations affect the behavior of the physical quantities characterizing TBIs such as the spin Hall conductivity, since most of the previous studies have focused on the bulk gap and the edge states at zero temperature. This issue is particularly important to discuss finite-temperature properties because the strong correlation should largely change the relevant energy scales of the system, bringing about a strong renormalization of the physical quantities. 

The purpose of this paper is twofold. We first elucidate how the correlation effects modify electronic properties in the TBI phase at finite temperatures with particular emphasis on the spin Hall conductivity. The deviation of the spin Hall conductivity from the quantized value at finite temperatures is clarified in terms of electron correlations. To this end, we employ a generalized Bernevig-Hughes-Zhang (BHZ) model having local electron correlations, and treat it  with dynamical mean field theory (DMFT) \cite{DMFT_Georges,DMFT_Metzner,DMFT_Muller-Hartmann}. 
It will be revealed that the spectral gap of the TBI is generically reduced by the correlation effect, as a result of band inversion,
leading to the strong temperature dependence of the spin Hall conductivity.
We then discuss how the TBI changes into the MI that could emerge in a strongly correlated region.  We will show that a first-order transition occurs between the TBI and the Mott insulator (MI), as far as the nonmagnetic Mott phase is concerned,
and thus a gap closing does not occur at the TBI-Ml transition.
The magnetic instability is also discussed.

\section{Model and Method}

\begin{figure}[!h]
\begin{minipage}{0.5\hsize}
\begin{center}
\includegraphics[width=40mm,clip]{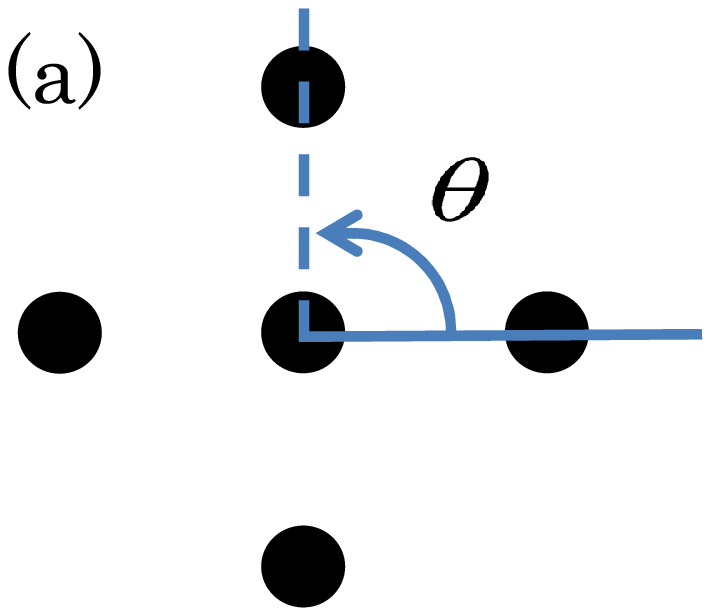}
\end{center}
\end{minipage}
\begin{minipage}{0.3\hsize}
\begin{center}
\includegraphics[width=35mm,clip]{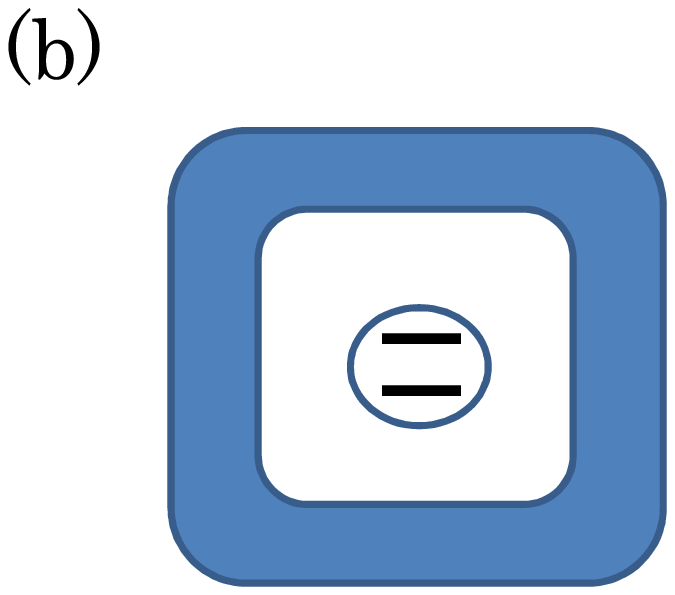}
\end{center}
\end{minipage}
\caption{
(a) Sketch of the BHZ model on a square lattice. The hopping integral between orbitals depends on $\theta$ which defines the angle between hopping direction and $x$-axis.
(b) Sketch of the DMFT. The lattice model is mapped to an effective impurity model.
}
\label{fig:model}
\end{figure}

We extend the BHZ model to include on-site Coulomb interactions. The Hamiltonian reads
\begin{eqnarray}
 H&=& H_{BHZ} +U\sum_{i,\alpha} n_{i,\alpha,\uparrow} n_{i,\alpha,\downarrow}\\
  H_{BHZ}&=& \sum_{i,\alpha,\sigma} \epsilon_\alpha n_{i,\alpha,\sigma} -\sum_{\langle i,j\rangle ,\sigma} \hat{c}^{\dagger}_{i,\alpha,\sigma} \hat{t_\sigma}_{\alpha,\alpha'} \hat{c}_{j,\alpha',\sigma},\\
-\hat{t_\sigma} &=&\left(
\begin{array}{cc}
 -t_{1}& it_{sp}e^{i \theta \sigma}  \\
it_{sp}e^{-i \theta \sigma} & t_{2}
\end{array}
\right),
\end{eqnarray}
where $n_{i,\alpha,\sigma}=c^{\dagger}_{i,\alpha,\sigma}c_{i,\alpha,\sigma}$. The operator $c^{\dagger}_{i,\alpha,\sigma}(c_{i,\alpha,\sigma})$ creates (annihilates) an electron at site $i$ and orbital $\alpha=1,2$ in spin state $\sigma=\uparrow, \downarrow$. Off-diagonal elements of the hopping matrix $\hat{t}$ drive the system into a non-trivial band insulator in the non-interacting case. The angle $\theta$ in the hopping matrix $\hat{t}$ specifies the direction of hopping (see Fig. \ref{fig:model}(a)).
We analyze the system with DMFT, which  treats local correlations exactly and is suitable for the systematic calculation for arbitrary strength of the Coulomb interaction. In DMFT, the original lattice problem is mapped onto an effective impurity model, which is solved self-consistently. 
The self-consistent equation for a paramagnetic phase is given by
\begin{eqnarray}
 \hat{g}_{\sigma}^{-1}(i\omega)&=& [\sum_{{\bf k}}\frac{1}{i\omega \mathbb{I}-\hat{h}_{\sigma}({\bf k})-\hat{\Sigma}_{\sigma}(i\omega)} ]^{-1} +\hat{\Sigma}_{\sigma}(i\omega), 
\end{eqnarray}
where $\hat{h}_{\sigma}({\bf k})$ is the Fourier transform of the hopping matrix. The self-energy of the lattice Green's function $\hat{\Sigma_{\sigma}}(i\omega)$ can be computed from the Green's function $\hat{g}(i\omega)$ for the effective impurity model.

To solve the impurity problem, the continuous-time quantum Monte Carlo method (CT-QMC) is employed here. In this method, the full Green's function is obtained from the following equations \cite{CT-QMC_segment_Werner,CT-QMC_kink_Werner,CT-QMC_kink_Haule}:
\begin{eqnarray}
 \hspace{-1cm}G_{imp}(\tau-\tau')&=&-\langle T c(\tau) c^\dagger(\tau') \rangle, \\
 \langle A \rangle  &=&\frac{1}{\mathrm{Z}_{loc}} \int D[c]D[c^\dagger]  \hat{A} \nonumber \\ 
 &&e^{-\int d{\tau_1} H_{loc}(\tau_1)+\int d\tau_1 d\tau_2 c^\dagger (\tau_1) \hat{F}(\tau_1-\tau_2) c(\tau_2)}, \label{eq:gimp_ct-qmc} \\
\hat{F}_{\alpha,\alpha'}(i\omega) &=& [i\omega -\epsilon_\alpha +g^{-1}(-i\omega)]_{\alpha,\alpha} \delta_{\alpha,\alpha'}, \\
H_{loc} &=& \sum_{\alpha,\sigma} (\epsilon_\alpha-\mu) n_{\alpha,\sigma} +Un_{\alpha,\uparrow}n_{\alpha,\downarrow}, 
\end{eqnarray}
where $\mathrm{Z}_{loc}$ is the partition function for the effective impurity model. In this method, Eq. (\ref{eq:gimp_ct-qmc}) is expanded with respect to the hybridization functions $\hat{F}_{\alpha,\alpha'}$, and  the resulting integral is evaluated with the Monte Carlo method. Absence of the Trotter decomposition makes it easier to access the low temperature region. For simplicity, we study the particle-hole symmetric case and choose the model parameters as $t_{1}=t_{2}=t$, $t_{sp}=0.5t$ and $\epsilon_1 (\epsilon_2)=-t (t)$. The hopping integral $t$ is chosen as the energy unit.\\

\section{Results}
 
\subsection{spin Hall conductivity}

\begin{figure}[!h]
\begin{center}
\includegraphics[width=80mm,clip]{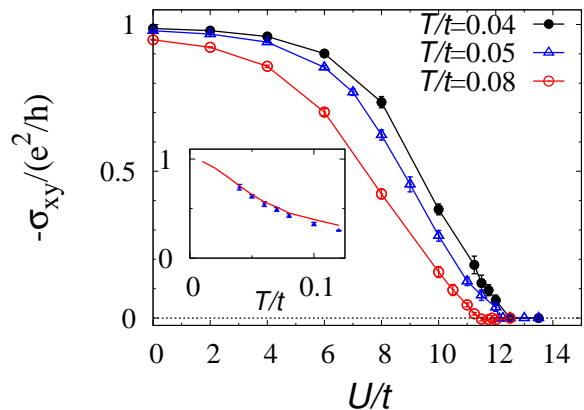}
\end{center}
\caption{(Color online)
The spin Hall conductivity as a function of the interaction strength. Solid circles, open triangles, and open circles denote the conductivity at $T=0.04$, $0.05$, and $0.08$ respectively.
Inset: spin Hall conductivity as a function of temperature. Open circles show the data at $U=8$. The solid red line shows the data calculated with the renormalized parameters which are obtained by low-energy expansion at $U=8$.
}
\label{fig: current}
\end{figure}
Let us first discuss the spin Hall conductivity, which characterizes the topological nature of the system. The spin Hall conductivity is calculated with the following formula,
\begin{eqnarray}
 \sigma_{xy}^{SH} &=& \frac{e^2}{\hbar}\mathrm{Im} \frac{\partial }{\partial \omega} K^R(\omega +i\delta),\\
 K(i\omega) &=& -\frac{1}{V\beta} \sum_{{\bf k}, i\nu,\sigma} \frac{\mathrm{sign}(\sigma)}{2} \mathrm{tr}[\frac{\partial \hat{h}_\sigma}{\partial k_y}({\bf k}) \nonumber \\
 && \hat{G}_{\sigma}({\bf k}, i\omega+i\nu) \frac{\partial \hat{h}_\sigma}{\partial k_x}({\bf k}) \hat{G}_{\sigma}({\bf k}, i\nu)], \label{eq: curr}
\end{eqnarray}
where $V$ denotes the volume of the system. Here, some comments are in order on the calculation of the conductivity. As well as in the non-interacting case, the conductivity should be a multiple of $e^2/h$ at zero temperature, which is checked as follows. Taking into account that the conductivity is antisymmetric tensor and that the vertex part is connected through the Ward identity ($\hat{\Lambda}_\mu (p,p)=\frac{\partial }{\partial p_{\mu}} \hat{G}^{-1}(p)$), we may represent the conductivity as the following equation:
\begin{eqnarray}
 \sigma^{SH}_{xy} &=& -\frac{e^2}{(2\pi)\hbar}N,\\
 N&=&\frac{\epsilon_{\mu\nu\rho}}{48\pi^2}\int d^3 p\sum_{\sigma} \mathrm{sign}(\sigma) \mathrm{tr}[ \frac{\partial \hat{G}_{\sigma}^{-1}(p)}{\partial p_{\mu}} \nonumber\\
 &&\hat{G}_\sigma(p) \frac{\partial \hat{G}_{\sigma}^{-1}(p) }{\partial p_{\nu} } \hat{G}_\sigma(p) \frac{\partial \hat{G}^{-1}_\sigma(p) }{\partial p_{\rho}} \hat{G}_{\sigma}(p)   ],
\end{eqnarray}
where the notation $p=(i\omega,{\bf p})$ is used.  
$N$ is the first Chern number expressed in terms of the Green's function \cite{volovik}.
Therefore, even in the correlated system, the conductivity should be a multiple of $e^2/h$ at zero temperature \cite{ch_num_Ishikawa,ch_num_Haldane}.
Besides, in the DMFT framework, the self-energy has no momentum dependence, and we can replace the vertex part with ${\partial \hat{h}_\sigma}/{\partial k_{\mu}}$. Thus, the conductivity, calculated via eq. (\ref {eq: curr}), is exact within the DMFT framework.
The analytic continuation is approximately done by replacing $\mathrm{Im}$ ${\partial K^R(\omega+i\delta)}/{\partial \omega }$ with $\mathrm{Im}$ ${\partial K(i\omega)}/{\partial i\omega }$. 

Note that in the above expression of $\sigma_{xy}^{SH}$, there are no current vertex corrections from quasiparticle interactions
which are related to the backflow, and normally exist in the expressions of transport coefficients.
This is because that the current vertex corrections are associated with current dissipations raised by quasiparticle scatterings,
and thus they do not appear in the expression of the dissipationless Hall conductivity with which we are concerned here.
Therefore, the above formula for the spin Hall conductivity is exact even for interacting electron systems.

We now turn to the numerical results. The calculated spin Hall conductivity is plotted in Fig. \ref{fig: current} as a function of the interaction strength at several temperatures. It is seen that the conductivity in the non-interacting case takes values close to unity in units of $e^2/h$ with little temperature dependence, implying that the system is indeed in the TBI phase.  When the interaction is introduced, the conductivity gradually decreases and shows a strong temperature dependence with increasing $U$. It eventually vanishes in the large interaction region.  Note that the conductivity in the TBI region is expected to reach unity at zero temperature. Actually, as seen in the inset of this figure, the conductivity for $U/t=8$  (open triangles) increases toward unity with decreasing temperature. We can confirm this tendency by focusing on the low-energy properties of the self-energy due to the Coulomb interaction; the computed values exhibit the almost identical behavior to the solid red line calculated with the renormalized parameters, which are obtained by the low-energy expansion of the self-energy as $\hat{\Sigma}_{\sigma}(\omega)=\hat{\Sigma}_{\sigma}(0) +\frac{\partial \hat{\Sigma}_{\sigma}}{\partial \omega}(0)\omega+O(\omega^2)$. At $T\lesssim 0.05$, no strong temperature dependence of the renormalized parameters is observed. We thus conclude that the spin Hall conductivity should reach unity at zero temperature and therefore the TBI should be stable for $U/t \lesssim 12$. 

It is instructive to point out that the above characteristic temperature/interaction dependence of the conductivity is not specific to the model parameters employed here, but more generic for correlated TBIs. As will be discussed momentarily below, the renormalization of the charge gap occurs due to interactions, which may increase the effective temperature with respect to the renormalized gap. This tendency is expected for the TBIs, so that the decrease of the conductivity with increasing temperature/interaction should be generally observed in TBIs.

\begin{figure}[!h]
\begin{minipage}{1\hsize}
\begin{center}
\includegraphics[width=85mm,clip]{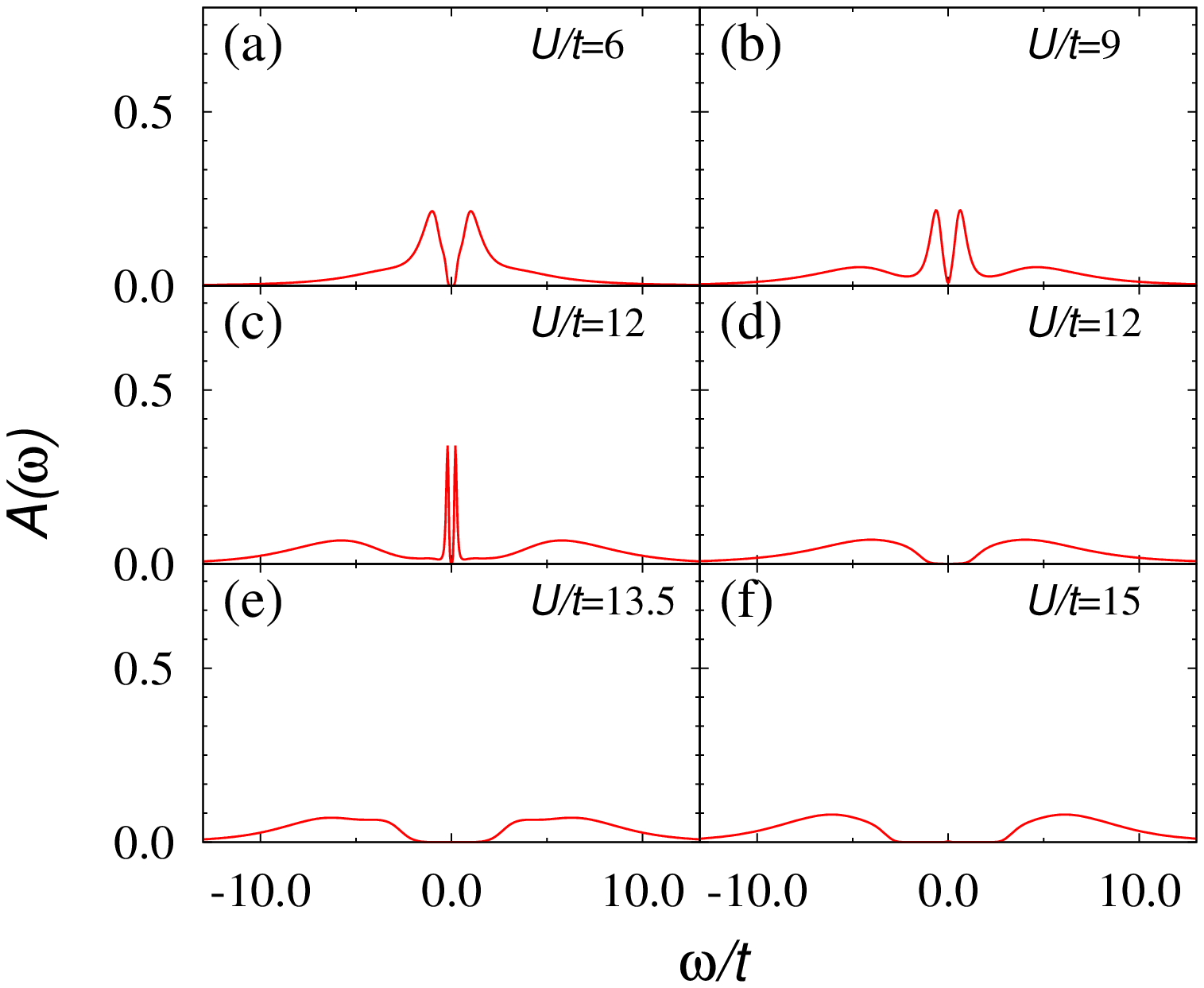}
\end{center}
\end{minipage}
\begin{minipage}{1\hsize}
\begin{center}
\includegraphics[width=85mm,clip]{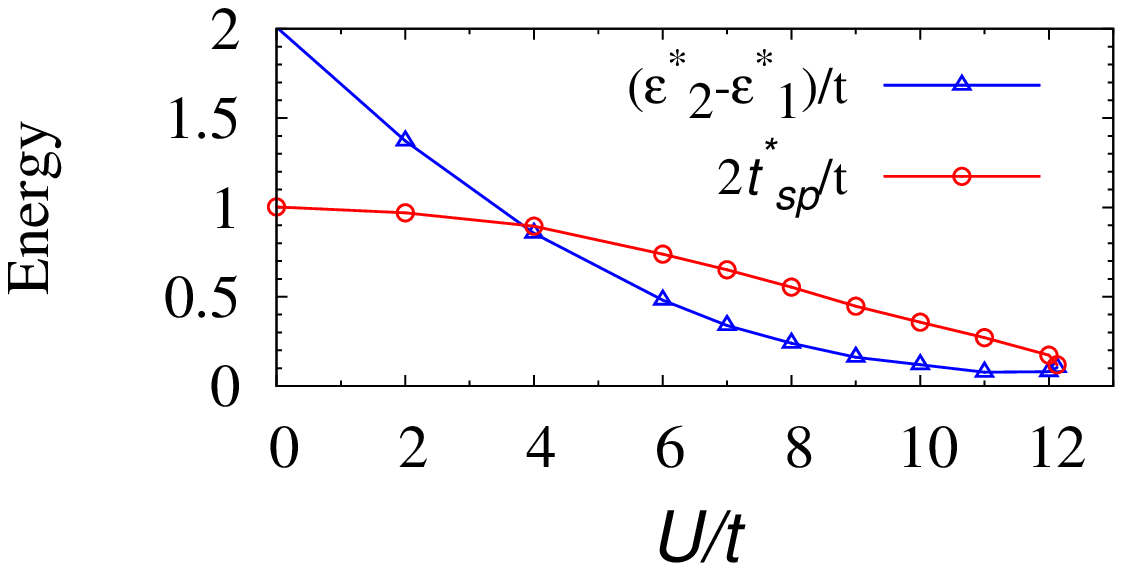}
\end{center}
\end{minipage}
\caption{(Color online)
Upper panel: Local density of states for each value of $U$. Figs. (a), (b) and (c) correspond to the TBI phase, while Figs. (d), (e) and (f) correspond to the MI phase, respectively. Note that in $11.5 <  U/t<12.125 $, a hysteresis is found. Lower panel: Renormalized energy splitting between the two orbitals.
}
\label{fig:DOS}
\end{figure}

\begin{figure}[!h]
\hspace{-3cm}
\begin{minipage}{0.4\hsize}
\begin{center}
\includegraphics[width=50mm,clip]{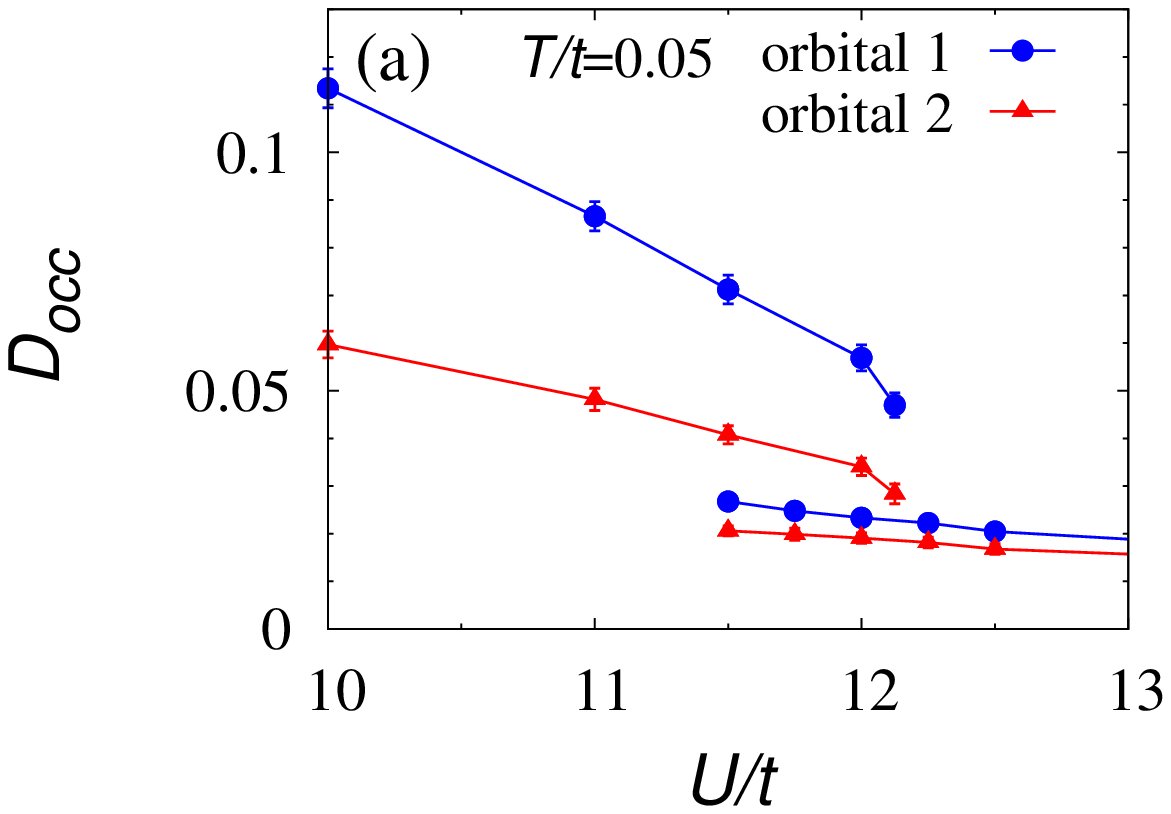}
\end{center}
\end{minipage}
\hspace{1cm}
\begin{minipage}{0.4\hsize}
\begin{center}
\includegraphics[width=55mm,clip]{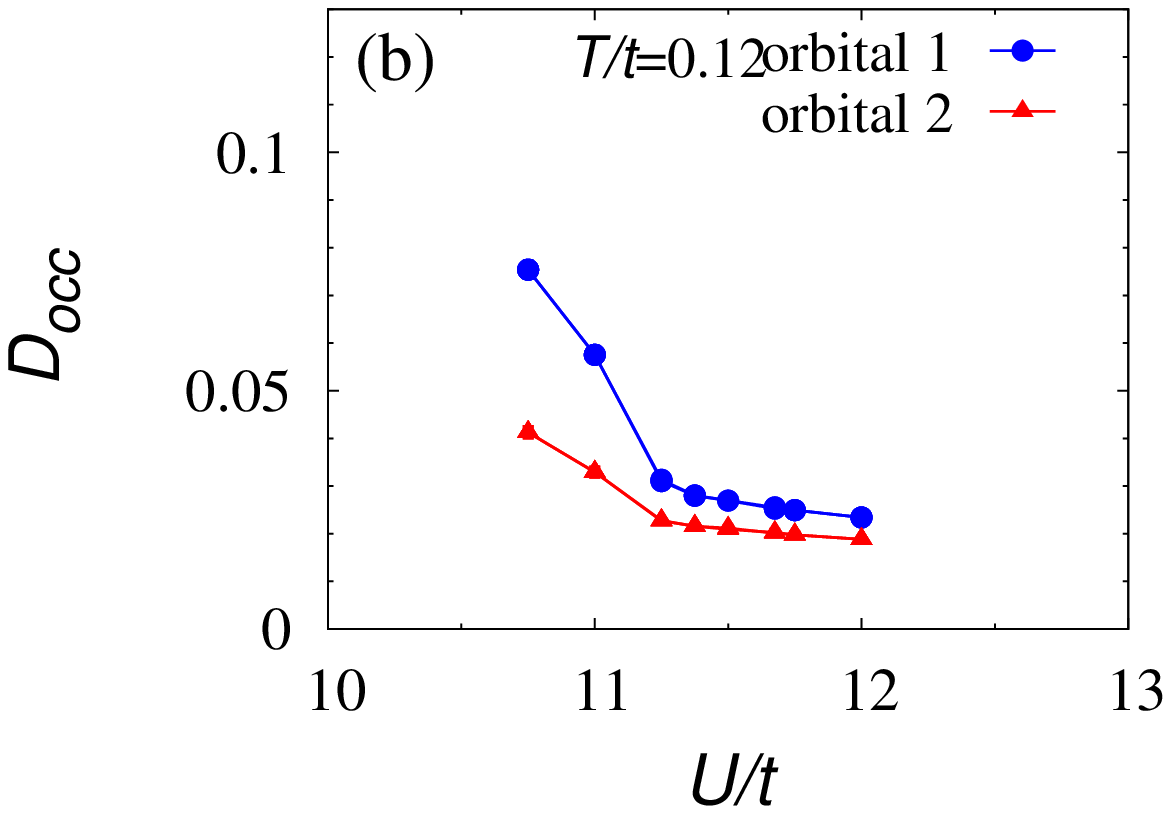}
\end{center}
\end{minipage}
\caption{(Color online)
Double occupancy of each orbital as a function of interaction: (a) $T/t$=0.05 (b) $T/t$=0.12. Note that at $T/t$=0.12, no hysteresis is found.
}
\label{fig:correlation}
\end{figure}

In order to clarify the origin of the strong-temperature dependence of the conductivity, let us now look at the one-particle spectral function. The spectral function ($A(\omega)=-\frac{1}{2\pi}\mathrm{Im}\mathrm{Tr}G(\omega+i\delta)$) is calculated at $T/t=0.05$. The analytic continuation procedure to convert the Matsubara frequency to the real frequency is done numerically with the maximum entropy method (MEM). The results obtained for several choices of interaction strength $U$ are plotted in the upper panel of Fig. \ref{fig:DOS}. In this figure, we can see the reduction of the gap and the evolution of Hubbard peaks with increasing $U$, which is consistent with the previous studies on the Kane-Mele-Hubbard model \cite{AFvsTBI_DQMC_Hohenadler11,AFvsTBI_VCA_Yu11,AFvsTBI_CDMFTWu11}. In the large $U$ region, the renormalized peaks in the low-energy region eventually disappear, implying that a quantum phase transition occurs between the TBI and the MI.

We note here that in generic band insulators the spectral gap is not necessarily reduced via the renormalization. Sometimes it is even enhanced by the interaction. Actually, according to Ref. \cite{renor-gap}, if the system has a local hybridization, the off-diagonal self-energy increases the local hybridization and hence enhances the gap.  In this sense, whether the gap is enhanced or reduced depends on the model systems employed. However, a remarkable point is that the gap for the TBIs should be always renormalized to be smaller due to electron correlations.  This intriguing property is a consequence of the band inversion inherent in the TBIs. Note that the TBIs have the gap resulting from the SOIs. If the interaction is switched on, the SOIs should be renormalized because the SOIs contribute to the kinetic energy. This renormalization gives the reduction of the spectral gap, as shown in Fig. \ref{fig:DOS}.  Although the renormalization of the energy splitting $\epsilon^*_2-\epsilon^*_1$, shown in the lower panel of Fig.\ref{fig:DOS}, seems to cause the gap renormalization, this renormalization should be irrelevant since it is essentially absorbed in the energy shift due to the band inversion in the TBIs. Therefore, we can say that the reduction of the gap  due to the interaction and the resulting strong-temperature dependence of the spin Hall conductivity are interesting generic features common to the TBIs.  
This result also implies that 
strong electron correlation is harmful to
the topological stability of the quantum spin Hall effect.

\subsection{Mott transition}

We now discuss what happens in the strong correlation region. In  Fig. \ref{fig:DOS}, we have already observed the disappearance of the renormalized peak structure for the TBI in the spectral function, but the nature of the transition is not clear yet. To address this issue, we compute the double occupancy of electrons at each site, which is a key quantity to characterize the Mott transition. In Fig. \ref{fig:correlation}, the double occupancy of each orbital is plotted as a function of the interaction strength. It is clearly seen in Fig. \ref{fig:correlation} (a) that the double occupancy exhibits a jump and a hysteresis around $U/t \sim 12$. This behavior indicates that the TBI changes into the MI via a first-order transition. Therefore, in the upper panel of Fig. \ref{fig:DOS}, the coherence peaks in the low energy region vanish suddenly, and the system acquires the Hubbard gap  in the MI phase. This is contrasted to the non-interacting case, where the gap closing occurs continuously when the TBI changes into a trivial insulator. In the MI phase, the gap size is of the order of $U$, so that the temperature effect is not important in this region. As seen in Fig. \ref{fig: current}, the spin Hall conductivity is almost zero, which confirms that the MI belongs to a trivial phase. 

In Fig. \ref{fig:correlation} (b), the results at higher temperatures are shown. It is seen that the first-order transition is now changed into a crossover where the jump and hysteresis are smeared. Note that this kind of crossover behavior is usually accompanied by the first-order Mott transitions at higher temperatures.

\begin{figure}[!h]
\begin{minipage}{0.3\hsize}
\begin{center}
\includegraphics[width=\hsize,clip]{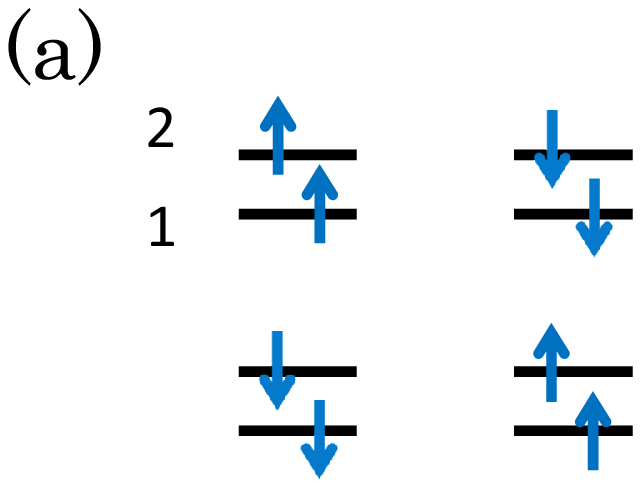}
\end{center}
\end{minipage}
\begin{minipage}{0.65\hsize}
\begin{center}
\includegraphics[width=\hsize,clip]{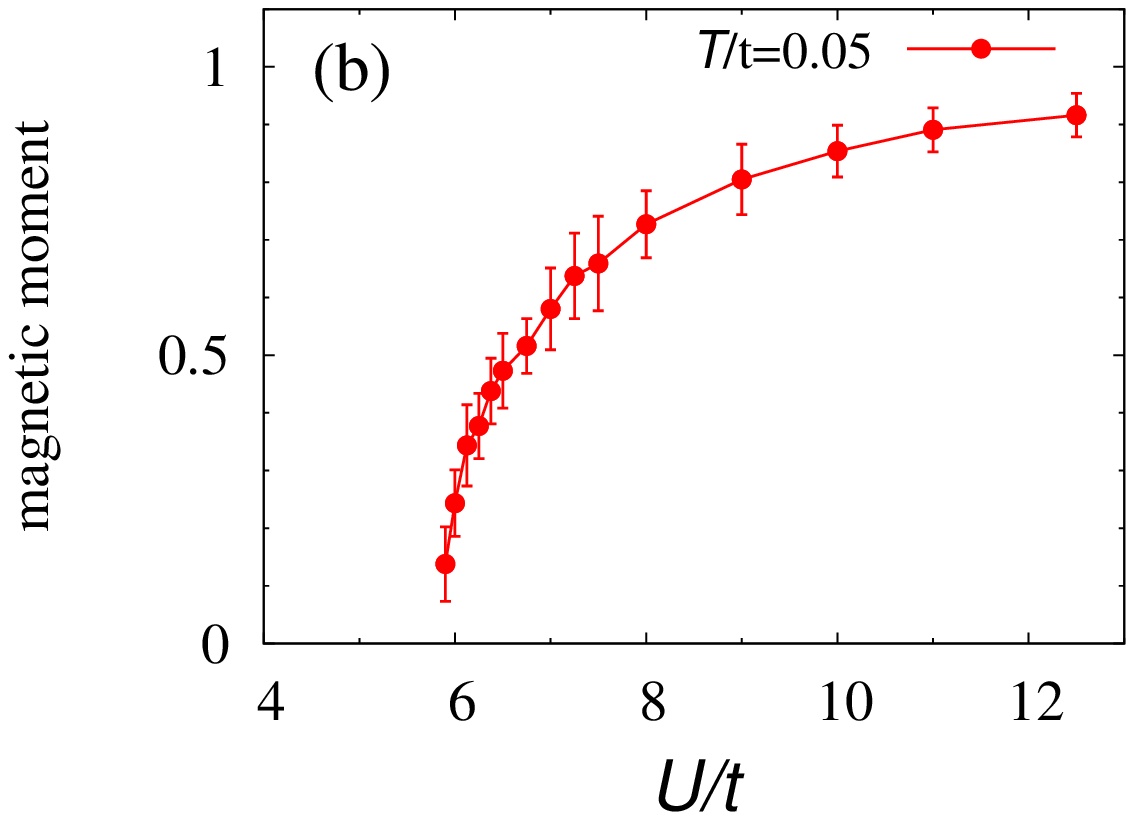}
\end{center}
\end{minipage}
\caption{
(a): Sketch of the spin configuration in the AF phase. Electron spins are denoted by the blue arrows. The numbers 1, 2 specify the orbital indices. (b): Magnetic moment as a function of the interaction at $T/t=0.05$.
}
\label{fig:AF_order}
\end{figure}

\subsection{magnetic instability}

\begin{figure}[!h]
\begin{center}
\includegraphics[width=80mm,clip]{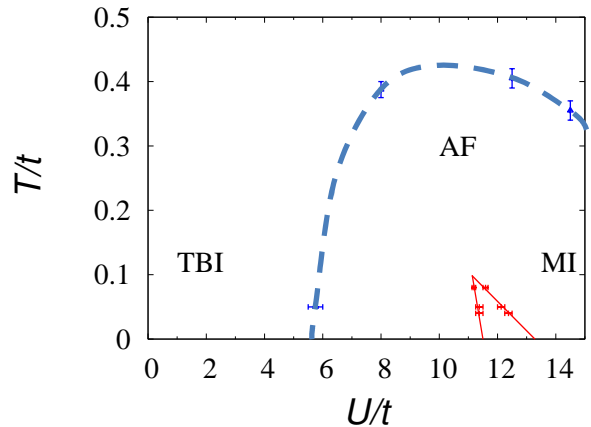}
\end{center}
\caption{(Color online)
The temperature vs. interaction phase diagram. The dotted blue line denotes the second order transition between the TBI phase and the AF phase. If the AF phase is suppressed, a first-order Mott transition could be observed (see text). Solid red lines denote the Mott transition line, where the coexistence phase accompanied by the hysteresis is found in the region surrounded by red lines.
}
\label{fig: phase}
\end{figure}

We have so far  restricted ourselves to a paramagnetic MI phase. Note that in the DMFT framework, the assumption of the paramagnetic MI phase corresponds to implicit introduction of frustration, which is not really included in the present model (see the discussions below).  Now, let us turn to the magnetic instability of the system. From the estimation of the exchange interactions with a perturbation theory in hopping, it is found that the off-diagonal elements of the hopping matrix $\hat{t}$ provide the exchange interactions which depend on spin components; the $z$-component of the exchange interaction is antiferromagnetic (AF), while the in-plane components are ferromagnetic (AF) between neighbors in $y$- ($x$-) direction. This results from the phase factor $\theta$ and spin dependency of the hopping matrix. From the above considerations, the spin configuration of the AF phase is expected to be the one shown in Fig. \ref{fig:AF_order}. This long-range AF ordered phase is treated with the sublattice method in the framework of DMFT. The order parameter, which is defined by $m=\sum_{\alpha}(n_{\alpha,\uparrow}-n_{\alpha,\downarrow})/2$, is plotted in Fig.\ref{fig:AF_order}(b). In this figure, we can find a second-order quantum phase transition. By performing similar calculations for various sets of parameters, we end up with the phase diagram shown in Fig. \ref{fig: phase}. It is seen that the paramagnetic Mott transition is now masked by the AF phase. This kind of behavior is commonly observed in the DMFT treatment for the MI transition. Therefore, we naturally expect the transition from the TBI to the AF insulator for our generalized BHZ model. 
The above results imply that the spectral gap which persists up to $U \sim 12t$ in the paramagnetic case does not close at the magnetic transition at $U\sim 5t$.
This behavior is similar to the result obtained for the Kane-Mele-Hubbard model in Ref.\cite{AFvsTBI_DQMC_Hohenadler11},
though another research work reports that the gap closing occurs at the transition \cite{AFvsTBI_VCA_Yu11}.

If the system is geometrically frustrated, however, the paramagnetic MI phase mentioned in the previous subsection might be stabilized. Here, we make a comment on whether we can have a chance to expect the Mott transition in such frustrated systems. A frustrated system in the BHZ model can be realized if the next nearest neighbor (NNN) hopping is taken into account in the square lattice. We have confirmed that in the particle-hole symmetric case, the system is in the TBI phase at $U=0$ if the NNN hopping meets the condition: $\epsilon_2-\epsilon_1<4(t-t')$, where $t'$ represents the NNN hopping. Therefore, we could have a quantum phase transition between the TBI and the nonmagnetic MI discussed here, starting from the frustrated BHZ model. This problem is now under consideration. 

In this connection, we wish to mention that there are some other candidates for TBIs which have geometrically frustrated lattice structure. A certain extension of a kagom\'e lattice, which has a non-trivial topological band structure \cite{Kagome_TBI_Franz09}, and pyrochlore iridium oxides such as $\mathrm{A_2Ir_2O_7}$ $(\mathrm{A=Pr, Eu})$ \cite{TMI_LBalents09,Kargarian11} are proposed as candidates of frustrated TBIs. In these frustrated systems, we may expect that the first-order quantum phase transition should occur between the TBI and the MI, if we could realize the TBI in the weak coupling region.

Before closing this section, we make a brief comment on a possible topological aspect of the antiferromagnetic phase.  It was previously proposed that a topologically nontrivial phase could appear in the antiferromagnetic region \cite{AFTI_Moore10}. 
However, we have not found the non-zero spin Hall conductivity in this phase within the present approach. Detailed study on this issue is interesting, which is to be done in the future work.

\section{Summary}

In this paper, we have studied the effects of the local Coulomb interaction on the TBI by applying DMFT and CT-QMC to a generalized BHZ model. We have calculated the spin Hall conductivity and found that the strong temperature dependence of the spin-Hall conductivity results from the renormalization of the gap due to electron correlations. We have pointed out that the reduction of the gap via the renormalization is a generic feature common to the TBIs, and is raised by a band inversion structure characteristic of TBIs. 
We have also shown that the TBI-MI transition is of first-order if we restrict ourselves to a paramagnetic Mott phase. This condition corresponds to taking into account frustration in the DMFT framework. 
If the magnetic instability is considered in our generalized BHZ model, the Mott transition is masked and the transition becomes of second order. 
Our results imply that a gap closing process does not occur for both cases of the paramagnetic Mott transition and
and the magnetic transition.
We have also discussed possible situations in which the paramagnetic Mott transition would be observed for the systems with frustration. It is an interesting open issue to treat the Mott transition for such frustrated systems in topologically nontrivial conditions, such as Iridium oxides. The application of DMFT calculation to such systems is now under consideration.

\begin{acknowledgments}

A part of numerical computations was done at
the Supercomputer Center at ISSP, University of Tokyo
and also at YITP, Kyoto University. This work was
supported by KAKENHI (Nos. 21540359, 20102008, 21102510, 23540406, and 19052003), 
the Grant-in-Aid for the Global COE Programs"
The Next Generation of Physics, Spun from Universality
and Emergence " from MEXT of Japan. N.K.
is supported by JSPS through its FIRST ProgramD

\end{acknowledgments}



\begin{thebibliography}{99}

\bibitem{Hasan10}	M. Z. Hassan and C. L. Kane, Rev. Mod. Phys. {\bf 82}  3045-3067  (2010).
\bibitem{Qi10}	X. L. Qi and S. C. Zhang, Rev. Mod. Phys. {\bf 83}, 1057 (2011).

\bibitem{Bernevig06}	B. A. Bernevig and S. C. Zhang, Phys. Rev. Lett. {\bf 96}  106802 (2006).

\bibitem{Kane05_1}	C. L. Kane and E. J. Mele, Phys. Rev. Lett. {\bf 95} 146802   (2005).
\bibitem{Kane05_2}	C. L. Kane and E. J. Mele, Phys. Rev. Lett. {\bf 95} 226801   (2005).
\bibitem{Bernevig06_BHZ}	B. A. Bernevig, T. L. Hughes, and S. C. Zhang, Science {\bf 314}  1757  (2006).

\bibitem{3DTI_Fu07} L. Fu, C. L. Kane, and E. J. Mele, Phys. Rev. Lett. 98 106803 (2007).
\bibitem{TBI_3DFu07}	L. Fu and C. L. Kane, Phys. Rev. B {\bf 76} 045302 (2007).
\bibitem{Qi08}	X. L. Qi, T. L. Hughes and S. C. Zhang, Phys. Rev. B {\bf 78} 195424   (2008).

\bibitem{exp_2D-QW_MKonig} M. K\"onig, S. Wiedmann, C. Br\"une, A Roth, H. Buhmann, L. W. Molenkamp, X. L. Qi, and S. C. Zhang, Sience{\bf 318} 776 (2007).
\bibitem{exp_3D-bismuth_DHsieh} D. Hsieh, D. Qian, L. Wray, Y. Xia, Y.S. Hor, R. J. Cava, and M. Z. Hassan, Nature {\bf 452} 970 (2008).
\bibitem{exp_3D-bismuth_YXia} Y. Xia, D. Qian, D. Hsieh, L. Wray, A. Pal, H. Lin, A. Bansil, D. Grauer, Y. S. Hor, R. J. Cava, and M. Z. Hassan, Nat. Phys. {\bf 5} 398 (2009).
\bibitem{exp_3D-bismuth_YLChen} Y. L. Chen, J. G. Analytis, J.-H. Chu, Z. K. Liu, S.-K.Mo, X. L. Qi, H. J. Zhang, D. H. Lu, X. Dai, Z. Fang, S. C. Zhang, I. R. Fisher, Z. Hussain, and Z.-X. Shen, Science {\bf 325} 178 (2009).


\bibitem{NaIrO_Nagaosa09} A. Shitade, H. Katsura, J. Kunes, X.-L. Qi, S. C. Zhang, and N. Nagaosa, Phys. Rev. Lett. {\bf 102} 256403 (2009).

\bibitem{TMI_LBalents09} D. A. Pesin and L. Balents, Nature Physics {\bf 6}, 376 - 381 (2010) .



\bibitem{Raghu08}		S. Raghu, X. L. Qi, C. Honerkamp, and S. C. Zhang, Phys. Rev. Lett. {\bf 100} 156401 (2008).
\bibitem{TM-Z_YZhang09}	Y. Zhang, Y. Ran, and A. Vishwanath, Phys. Rev. B {\bf 79} 245331 (2009).
\bibitem{Kurita11}	M. Kurita, Y. Yamaji, and M. Imada, J. Phys. Soc. Jpn. {\bf 80} 044708 (2011).
\bibitem{Fiete11} 	G. A. Fiete, V. Chua, X. Hu, M. Kargarian, R. Lundgren, A. R{\"u}egg, J. Wen and V. Zyuzin, arXiv: 1106.0013v1


\bibitem{Yamaji11}	Y. Yamaji and M. Imada, Phys. Rev. B {\bf 83} 205122 (2011).
\bibitem{AFvsTBI_DQMC_Hohenadler11}	M. Hohenadler, T. C. Lang, and F. F. Assaad. Phys. Rev. Lett.  {\bf 106} 100403 (2011).
\bibitem{AFvsTBI_DQMC_edge_early_Zheng10}	D. Zheng, C. Wu, and G-M Zhang, arXiv: 1011.5858v2 (2010).
\bibitem{AFvsTBI_VCA_Yu11}	S. L. Yu, X. C. Xie, and J. X. Li, Phys. Rev. Lett. {\bf 107} 010401 (2011).
\bibitem{AFvsTBI_CDMFTWu11}	W. Wu , S. Rachel, W-M Liu, and K. L. Hur, arXiv:1106.0943v1. (2011)
\bibitem{Rachel10}	S. Rachel and K. LeHur, Phys. Rev. B {\bf 82} 075106   (2010).

\bibitem{Varney10}	C. N. Varney, K. Sun, M. Rigol, and V. Galitski, Phys. Rev. B {\bf 82} 115125 (2010).
\bibitem{Wang10}	L. Wang, H. Shi, S. Zhang, X. Wang, X Dai and X. C. Xie, arXiv. 1012.5163v1 (2011).


\bibitem{DMFT_Georges} A. Georges, G. Kotliar, W. Krauth and M. J. Rozenberg,  Rev. Mod. Phys. {\bf 68} 13-125 (1996).
\bibitem{DMFT_Metzner} W. Metzner and D. Vollhardt, Phys. Rev. Lett. {\bf 62} 324 (1989). 
\bibitem{DMFT_Muller-Hartmann} E. M. Hartmann, Z. Phys. B Condens. Matter {\bf 74} 507 (1989).



\bibitem{CT-QMC_segment_Werner} P. Werner, A. Comanac, L. de'Medici, M. Troyer and A. J. Millis, Phys. Rev. Lett. {\bf 97} 076405 (2006).
\bibitem{CT-QMC_kink_Werner} P. Werner and A. J. Millis, Phys. Rev. B. {\bf 74} 155107 (2006).
\bibitem{CT-QMC_kink_Haule} K. Haule, Phys. Rev. B. {\bf 75} 155113 (2007).

\bibitem{volovik} G. E. Volovik, {\it The Universe in a Helium Droplet} (Oxford, 2003).
\bibitem{ch_num_Ishikawa} K. Ishikawa and T Matsuyama, Nucl. Phys. B {\bf 280} 523-548 (1987).
\bibitem{ch_num_Haldane} F. D. M. Haldane, Phys. Rev. Lett. {\bf 93} 206602 (2004).

\bibitem{renor-gap} R. Sato, T. Ohashi, A. Koga and N. Kawakami, J. Phys. Soc. Jpn. {\bf 73} 1864-1869 (2004).


\bibitem{Kargarian11} M. Kargarian, J. Wen and G. A. Fiete, Phys. Rev. B {\bf 83} 165112 (2011).

\bibitem{Kagome_TBI_Franz09} H-M. Guo and M. Franz, Phys. Rev. B {\bf 80} 113102 (2009).
\bibitem{AFTI_Moore10}  R. S. K. Mong, A. M. Essin, and J. E. Moore , Phys. Rev. B {\bf 81} 245209 (2010).



\end{thebibliography}
\end{document}